\documentclass[12pt]{article}
\usepackage{pic02}
\usepackage{hyperref}
\usepackage{url}
\usepackage{graphicx}
\def\BaBar{\slshape{B{\small A}B{\small AR}}}
\def\ifmath#1{\relax\ifmmode #1\else $#1$\fi}
\def\ra  {\ifmath{\rightarrow}}           %
\def\Vud  {\ifmath{V_{\!ud}}}           %
\def\Vus  {\ifmath{V_{\!us}}}           %
\def\Vub  {\ifmath{V_{\!ub}}}           %
\def\Vcd  {\ifmath{V_{\!cd}}}           %
\def\Vcs  {\ifmath{V_{\!cs}}}           %
\def\Vcb  {\ifmath{V_{\!cb}}}           %
\def\Vtd  {\ifmath{V_{\!td}}}           %
\def\Vts  {\ifmath{V_{\!ts}}}           %
\def\Vtb  {\ifmath{V_{\!tb}}}           %
\def\Vij  {\ifmath{V_{\!ij}}}           %
\def\rhobar  {\ifmath{\bar{\rho}}}           %
\def\etabar  {\ifmath{\bar{\eta}}}           %

\def\ufs{\ifmath{\Upsilon(4\rm S)}}       
\def\Lom{\ifmath{\Lambda_{\rm QCD}/m_b}}           %
\def\Lbar{\ifmath{\bar{\Lambda}}}           %
\def\lone{\ifmath{\lambda_1}}           %
\def\ltwo{\ifmath{\lambda_2}}
\def\bra{\ifmath{\langle}}
\def\ket{\ifmath{\rangle}}
\def\MXtwo{\ifmath{\bra M_X^2 \ket}}
\def\plstarmin{\ifmath{p_{\ell, min}^*}}
\def\dmd{\ifmath{\Delta m_d}}

\def\mev {\ifmath{\rm \,MeV}}
\def\gev {\ifmath{\rm \,GeV}}

\def\ps {\ifmath{\rm \,ps}}

\def\eg  {{\it e.g.}}
\def\ie  {{\it i.e.}}
\def\etal{{\it et al.}}
\def\vs  {{\it vs}}

\def\etc  {{\rm etc.}}
\begin{document}

\title{\bf MEASUREMENTS OF CKM ELEMENTS \\ 
           AND \\
           THE UNITARITY TRIANGLE}
\author{
Helmut Marsiske        \\
{\em Stanford Linear Accelerator Center, Stanford University, Stanford, 
CA 94309}}
\maketitle

%
%
%
%
%
%
\vspace{4.5cm}
%

\baselineskip=14.5pt
\begin{abstract}
In this presentation, I review the status of selected 
Cabibbo-Kobayashi-Maskawa (CKM) matrix elements
and their role in the Unitarity Triangle (UT). 
Since this conference concluded, many new results have been
finalized and are included in the new world averages (WAs) 
in the 2002 edition of the PDG~\cite{PDGckm, PDGvcb, PDGvub}.
I will focus here on some outstanding issues in the measurements of 
the CKM elements in the third row and column.

    \begin{center}
\[    \left( \begin{array}{c} d' \\ s' \\ b' \end{array} \right) = 
                \left( \begin{array}{ccc}
	          \Vud& \Vus& \Vub \\
	          \Vcd& \Vcs& \Vcb \\
	          \Vtd& \Vts& \Vtb
	        \end{array} \right)
    \left( \begin{array}{c} d \\ s \\ b \end{array} \right)
\]
    \end{center}

\end{abstract}
\newpage

\baselineskip=17pt

\section{CKM, UTs, and Theoretical Tools}
In the Standard Model (SM)~\cite{PDGsm}, the charged-current 
electro-weak interactions amongst
three generations of quarks involve a complex, unitary $3\times 3$ 
matrix, $V$, known as the CKM matrix~\cite{PDGckm}. 
Its elements, \Vij, determine 
the relative (weak) couplings of up-type (\ie, $i = u, c, t$) 
and down-type (\ie, $j = d, s, b$) quarks to the $W$~boson. 
The matrix $V$ is determined by four independent parameters, one
of which can be used to describe CP violation (CPV). 
A parameterization developed by Wolfenstein emphasizes 
the hierarchy of elements by expanding them in powers of the sine 
of the Cabibbo angle: 
$\lambda = |\Vus| = \sin \theta_C \approx 0.22$. 
The other three parameters are labelled $A$, $\rho$, and $\eta$, 
with $\eta$ describing the level of CPV.
In the improved Wolfenstein parameterization~\cite{burasfleischer},
using
$\rhobar = \rho (1 - \frac{\lambda^2}{2})$
and
$\etabar = \eta (1 - \frac{\lambda^2}{2})$,
the matrix reads:

\begin{equation}
V =     \left( \begin{array}{ccc}
                    {
                    1-\frac{\lambda^2}{2}}&
                    {
                    \lambda}&
                    {
                    A\lambda^3(\rho-i\eta)} \\ 
                    {
                    -\lambda-iA^2\lambda^5\eta}&
                    {
                    1-\frac{\lambda^2}{2}}&
                    {
                    A\lambda^2} \\ 
                    {
                    A\lambda^3(1-\bar{\rho}-i\bar{\eta})}&
                    {
                    -A\lambda^2+A\lambda^4(\frac{1}{2}-\rho)-iA\lambda^4\eta}&
                    {
                    1-\frac{1}{2}A^2\lambda^4}
	          \end{array} \right)
         \ +\ {\cal{O}}(\lambda^6).
\label{eq:ckmwolf}
\end{equation}

The unitarity constraint results in six orthogonality equations
which can be expressed geometrically by six UTs in the complex
plane. All triangles have the same area, $\Delta$, which determines
the level of CPV:
$2 \Delta = J_{\rm CP} \approx A^2 \lambda^6 \eta$.
Most triangles are squashed, \ie, one side is very much smaller
than the others. However, there is one ``golden triangle''
with all sides of approximately equal size (\ie, ${\cal{O}}(\lambda^3)$),
determined by products of CKM elements that are experimentally 
most accessible:

\begin{equation}
\Vud\Vub^* + \Vcd\Vcb^* + \Vtd\Vtb^* = 0
\label{eq:UT}
\end{equation}

It is customary to re-scale its sides by
$|\Vcd\Vcb^*| = A\lambda^3$, 
resulting in a triangle in the complex plane with a unit-length baseline
on the real axis.
The UT in the $\bar{\rho}$, $\bar{\eta}$ plane
has sides of length:
\begin{eqnarray}
    R_u & = & \sqrt{ \rhobar^2 + \etabar^2 } 
          = (1-\frac{\lambda^2}{2}) \frac{1}{\lambda}
            \frac{|\Vub|}{|\Vcb|} 
          = \frac{|\Vud\Vub^*|}{|\Vcd\Vcb^*|}\,, \nonumber \\
    R_t & = & \sqrt{ (1 - \rhobar)^2 + \etabar^2 }
          = \frac{1}{\lambda}
            \frac{|\Vtd|}{|\Vcb|}
          = \frac{|\Vtd\Vtb^*|}{|\Vcd\Vcb^*|}\,. 
\label{eq:rurt}
\end{eqnarray}


In order to over-constrain the UT as a test of the SM, it is thus imperative
to measure as many of the angles as possible, and improve the measurements
of the relatively poorly known elements \Vub, \Vcb, and \Vtd\ which determine
the sides. (Note that \Vud\ and \Vus\ are know to about 0.1\% and 1\%, 
respectively.)

Six CKM elements, $V_{uq}$ and $V_{cq}$, $q = d, s,b$, can currently
be determined directly from tree-level processes. The remaining three
elements, $V_{tq}$, involve the top quark and are currently accessible 
only via loop processes in the $K$ and $B$~system (\ie, rare $K$ and 
$B$~decay, and $K^0/B^0_d/B^0_s$ mixing).
In the LHC/LC future, $V_{tq}$ and $V_{cq}$ will be measured at tree level
from top and $W$~decay, respectively, possibly with very high 
precision~\cite{lettsmaettig}.

A general problem in determining CKM matrix elements arises from the fact
that they enter into the charged currents between {\it quarks}, whereas
experiments observe initial and final states containing {\it hadrons}
and/or leptons.
The ``dressing'' of quarks makes for strong-interaction effects that 
are difficult to deal with, in particular since one is often in the 
non-perturbative regime of Quantum Chromodynamics (QCD). 
The problem is somewhat simplified for
semi-leptonic decays, because the (leptonic) $W$~decay is well-understood
and can be factored out.

The formalism to evaluate (weak) decay amplitudes of hadrons is that of 
the Operator Product Expansion (OPE) and Renormalization Group Evolution.
Using a scale, $\mu$, long- and short-distance contributions are separated
into perturbative (and thus calculable) Wilson coefficients,
$C_i(\mu)$,  and non-perturbative hadronic matrix elements, 
$\langle F| Q_i(\mu) |M \rangle$, written in terms of 
local operators generated by QCD and electroweak interactions:
\begin{equation}
            A(M\rightarrow F) = 
            \frac{G_F}{\sqrt{2}} \sum_{i} V_{CKM}^i C_i(\mu) 
            \langle F| Q_i(\mu) |M \rangle\,.
\label{eq:ope}
\end{equation}

The tool of choice to get at hadronic matrix elements,
in particular for {\it exclusive} final states,
is Lattice QCD (LQCD), which has recently made great strides towards
unquenched calculations, which could eventually yield quantifiable errors 
at the ${\cal{O}}$(1\%) level. 
Until then, other methods (\eg, $1/n_f$~expansion,
QCD sum rules, chiral perturbation theory, \etc) have to be used, the 
results of which are usually formulated in terms of (meson) decay constants
and form factors (FFs).

In case of the heavy-to-heavy 
$b\ra c$ transition,
Heavy Quark Effective Theory (HQET) can provide an absolute normalization
for FFs at $q^2 = (p_\ell + p_\nu)^2 = q_{max}^2$ and in the infinite 
quark mass limit, 
with corrections of the order of only 10\%. 
Unfortunately, no such luck for the heavy-to-light 
$b\ra u$ transition: the FF absolute normalization 
(and possibly the $q^2$ dependence)
must be calculated from the ground up.

{\it Inclusive} (semi-leptonic) decays of $B$~mesons turn out to be 
particularly favorable, for their treatment in the framework of Heavy 
Quark Expansion (HQE) allows a well-defined expansion of the decay rate
in powers of $\alpha_s$ (perturbative part) and \Lom\ (non-perturbative
part): 

\begin{equation}
\Gamma(B \ra X\ell\nu) = \Gamma(b \ra q\ell\nu) 
                       + {\cal{O}}\left[ \alpha_s, (\Lom)^2 \right]\,, 
\label{eq:GBXellnu}
\end{equation}

\noindent
where the first (and dominant) term represents the well-known 
quark spectator model rate (containing $|V_{qb}|^2$). 
Non-perturbative corrections are suppressed by at least two powers of
\Lom\ and can be expressed in terms of a small number of measurable 
parameters. 
This approach is valid only under the {\it ab initio} assumption 
of quark-hadron duality, which needs experimental verification. 
As has been advocated in Ref.~\cite{HF9bauer}, the semi-leptonic
data itself can be used to test for duality violations by checking
consistency among a variety of OPE-calculated variables.

\section{\Vcb~\cite{PDGvcb}}

$|\Vcb|$ normalizes the baseline of the UT and determines the 
Wolfenstein parameter $A$ (together with $\lambda = |\Vus|$, which is
very precisely known).

\subsection{From Exclusive Semi-leptonic $B$ Decays}

$|\Vcb|$ can be determined from the decay $B\rightarrow D^*\ell\nu$
by measuring the differential decay rate:
\begin{equation}
                  \frac{d\Gamma}{dw} \propto (F(w) |\Vcb|)^2\,,\
{\rm with}\
  w = \frac{m_B^2 + m_{D^*}^2 - (p_\ell + p_\nu)^2}{2m_B m_{D^*}}\,,
\label{eq:BDstarellnu}
\end{equation}

\noindent
and extrapolating the FF $F(w)$ to the kinematic limit $w = 1$, where
HQET
provides an absolute normalization
in the form of the Isgur-Wise function:
$F(w) = \eta_A \xi(w)$.
In the limit of infinite quark masses,  
$\xi(1) \equiv 1$.
Finite-mass and perturbative QED/QCD corrections are subsumed into the 
$\eta_A$ factor, yielding $F(1) = \eta_A = 0.91 \pm 0.04$.
This value for $F(1)$ results from a combination of quark model, OPE sum 
rule, and (quenched) LQCD calculations. It is promising to note that the
LQCD calculation can provide a rather detailed breakdown of the errors,
with the single largest one being statistical in nature.

Issues affecting this type of measurement are the assumed shape of the FF,
parameterized in terms of its slope, $\rho^2$, at $w=1$, the relative vector
and axial vector contributions to the FF, and the treatment of feed down
background from hadronic systems heavier than the  $D^*$.
A particular problem for experiments running on the \ufs~resonance lies in 
understanding the detection efficiency for the slow charged pion in the
$D^{*\pm}\ra \pi^\pm D^0$ decay. 
An important cross check is a simultaneous
measurement of the $D^{*0}\ra \pi^0 D^0$ decay, so far performed only
by CLEO.

All these issues need to be tackled in order to improve the measurement
beyond the current 5.2\% error (which is, however, dominated by the 
theoretical error on $F(1)$).
Moreover, the $B\rightarrow D^*\ell\nu$ and $B\rightarrow D\ell\nu$
data should be fit together to reduce the error on the FF slope,
since the slopes in both channels are related to each other in a 
calculable way.

The LEP \Vcb\ Working Group (WG)~\cite{LEPvcb} has combined the Belle, CLEO,
and LEP measurements of 
$F(1)|\Vcb|$ \vs\ $\rho^2$ 
-- note that those 
two quantities are highly correlated -- after adjusting for common inputs.
Most measurements and their errors are changed only slightly by this procedure,
except for the ALEPH result for which 
$F(1)|\Vcb|$ and $\rho^2$ increase by 1.5 $\sigma_{syst}^{old}$,
and $\sigma_{syst}^{new}$ is four times as large as the old one.
Unfortunately, the cited Ref.~\cite{LEPvcb} does not provide quite enough
information to reconstruct these calculations.
Despite these changes, the ALEPH and CLEO measurements remain rather far 
apart, resulting in a somewhat low confidence level (CL) of only 5\%
for the new WA. 
One should also point out that the $|\Vcb|$ value extracted from the
exclusive CLEO measurement by dividing out the above $F(1)$ comes out
rather high compared with the (presumably correlated) inclusive CLEO
measurement in Ref.~\cite{CLEOinclVcb}.
 

\subsection{From Inclusive Semi-leptonic $B$ Decays}

$|\Vcb|$ can also be extracted from the {\it inclusive} semi-leptonic
branching fraction (BF) ${\cal{B}}(B\ra X_c\ell\nu)$.
In order to distinguish prompt leptons (in $B\ra X_c\ell\nu$) from 
cascade leptons (in $B\ra X_c\ra X_{s/d}\,\ell\nu$), 
\ufs~experiments use a double-lepton technique:
a high-momentum lepton tags the opposite $B$, 
allowing to measure a signal-lepton down to
center-of-mass (COM) momenta of $p^* \approx 0.6\gev$
by exploiting charge and angular correlations.
Table~\ref{tab:inclBsl} shows the most current set of \ufs~measurements,
which are consistent with the LEP average.

\begin{table}
\centering
\caption{ \it Inclusive $B$ semi-leptonic BFs. 
In calculating the \ufs~average, the systematic errors have (wrongly) 
been assumed to be uncorrelated among experiments.}
\vskip 0.1 in
\begin{tabular}{|l|c|}\hline
Experiment & ${\cal{B}}(B\ra X \ell\nu)$ (\%)  \\ \hline\hline
BABAR prel.     & $10.87 \pm 0.18 \pm 0.30$                   \\ \hline
BELLE prel.     & $10.90 \pm 0.12 \pm 0.49$                   \\ \hline
CLEO 96    & $10.49 \pm 0.17 \pm 0.43$                   \\ \hline
CLEO 92    & $10.8  \pm 0.2  \pm 0.56$                   \\ \hline
ARGUS      & $ 9.7  \pm 0.5  \pm 0.4 $                   \\ \hline\hline
\ufs\ average & $10.66 \pm 0.08 \pm 0.18$                \\ \hline\hline
LEP average& $10.59 \pm 0.09 \pm 0.15 \pm 0.26$          \\ \hline\hline
\end{tabular}
\label{tab:inclBsl}
\end{table}

After subtracting the (small) $B\ra X_u\ell\nu$ contribution,
these BFs can be used to determine $|\Vcb|$, albeit with a 5.9\%
theoretical error\footnote{Note that the CL of this error is not 
particularly well defined: 
(not) following a suggestion in Ref.~\cite{Bigi9905} to combine
theoretical errors linearly, 
the LEP \Vcb\ WG decided to first scale them by a factor two before
combining them quadratically~\cite{CERNep2001050}.}.
As indicated in connection with Equation~\ref{eq:GBXellnu},
the theoretical error can be reduced 
by expressing the non-perturbative contributions to the semi-leptonic width
in terms of three parameters (at order $(\Lom)^2$): 
$\Lbar = m_B - m_b + \cdots\,$, \lone, and \ltwo,
all of which are {\it measurable} up to uncertainties of order $(\Lom)^3$.
The parameter \ltwo\ is readily determined from the 
$B^* - B$~mass splitting; the other two require more elaborate 
measurements.

Several experiments have embarked on a program to measure \Lbar\
(and thus $m_b$) and \lone\ from a variety of moments in semi-leptonic
and radiative $B$~decay data.
For example, CLEO has measured the first moment of the photon energy 
in $B\ra X_s\gamma$ to determine \Lbar, and the first moment of the
hadronic mass-squared, \MXtwo, recoiling against the di-leptons in 
$B\ra X_c\ell\nu$ to determine \lone. 
With these parameters as input, the above-mentioned (inclusive) 
semi-leptonic BFs have been used to determine $|\Vcb|$ with a 
2.6\% total error (and consistent with the exclusive result).

This all looks very promising and should improve even further, 
with preliminary CLEO~\cite{CLEOlepmom} and DELPHI~\cite{DELPHIlepmom} 
measurements 
of lepton energy moments in $B\ra X_c\ell\nu$ on their way.
One trouble spot has emerged, though, highlighted by the 
preliminary {\BaBar} measurement of \MXtwo\ \vs\ minimum lepton
COM momentum, \plstarmin~\cite{BABARichepMX}: while they obtain a result 
consistent with CLEO for the same \plstarmin cut of 1.5\gev, 
they find a \plstarmin\ dependence of \MXtwo\ that cannot 
be described by the CLEO values for \Lbar\ and \lone\ (but {\it can}
be described by OPE using a different set of parameter values).
This inconsistency suggests that some of the underlying assumptions
HQE is based on require further scrutiny.

\section{\Vub~\cite{PDGvub}}

The second side of the UT, $R_u$, is determined by $|\Vub|$;
see Equation~\ref{eq:rurt}.

\subsection{From Inclusive Semi-leptonic $B$~Decays}

Using the OPE, the inclusive semi-leptonic BF 
${\cal{B}}(B\ra X_u\ell\nu)$ can be related to $|\Vub|^2$ 
with an accuracy of $5-10$\%.
Unfortunately, experiments can only observe a small portion of
this decay: the overwhelming $B\ra X_c\ell\nu$ background forces them
to impose stringent analysis cuts, 
selecting a very restricted region
of phase space near kinematic boundaries 
(\ie, high lepton momentum, and/or low hadronic mass).
Under those circumstances, OPE cannot reliably predict the (partial) 
decay rate due to non-convergence.
The theorists way out of that is a ``twist expansion'' with the complete
series of terms re-summed into an incalculable structure function (light-cone
distribution function), describing the Fermi motion of the $b$~quark
inside the the $B$~meson. This structure/distribution function 
determines the decay rate
at leading order, with sub-leading twist corrections suppressed 
by (higher) powers of \Lom. 
Historically, it has been modeled in a somewhat {\it ad hoc} way, 
with the theoretical uncertainty in the fraction of events passing 
the analysis cuts estimated by varying the shape parameters.
This uncertainty can be removed in principle, though, because
the structure/distribution function is a property of the $B$~meson itself 
and can thus be determined (to leading order) in other processes, 
for example, $B\ra X_s\gamma$.
It has been argued in Ref.~\cite{LLW, BLM} that there is an uncontrollable 
uncertainty for $|\Vub|$, estimated at $\sim 15$\%,  
from transferring the $B\ra X_s\gamma$ structure/distribution function to
$B\ra X_u\ell\nu$
due to the sub-leading ${\cal{O}}$(\Lom) corrections.
In contrast, Ref.~\cite{Neubert} presents an actual 
calculation of the dominant 
source of power corrections, verifying the size of the effect, 
with a residual uncertainty for $|\Vub|$ of only a few percent.

The issue of sub-leading corrections needs to be resolved among the
theorists, because it affects the recent lepton-endpoint $|\Vub|$
measurements from CLEO~\cite{Bornheim} and {\BaBar}~\cite{BABARichepEll}
in one of two ways: either the theoretical error associated with this
needs to more than double (dominating then any other single error),
or the value of $|\Vub|$ needs to change by $10-15$\% (corresponding
to $2-3\sigma$ of the theoretical error associated with this previously).  
On top of all of this, one has to worry about the restrictive analyses
introducing quark-hadron duality violations, for which no estimate 
whatsoever exists currently.

One can nevertheless hold out hope for the future of (inclusive)
$|\Vub|$ measurements because of the enormous data samples becoming
available at the \ufs\ $B$~factories. 
The most promising analysis technique with hundreds of millions of $B$s
is that of fully reconstructing one of the two 
$B$s in the event, thus gaining complete control over the kinematics 
of the semi-leptonically decaying signal $B$. This allows rather precise
neutrino reconstruction, and charm suppression without kinematic cuts
using particle identification (\ie, no Kaon on the signal side). 
Furthermore, one can hope to 
use lepton momentum, hadronic mass, and di-lepton
mass simultaneously to isolate $B\ra X_u\ell\nu$ while retaining 
sufficient inclusiveness for OPE convergence and quark-hadron duality.

\subsection{From Exclusive Semi-leptonic $B$~Decays}

The very same $B$~reconstruction (Breco) technique will work beautifully
for exclusive decays, \ie, $B\ra \pi/\eta/\rho/\omega\, \ell\nu$,
providing very clean signals, with signal-to-background (S/B) ratios
$>>\,1$. Already with the data samples Belle and {\BaBar} have in hand,
the Breco technique is favored over the conventional analysis pioneered
by CLEO. 

The main problem for extracting $|\Vub|$ from these $B$~decay channels
is again theoretical in nature: a transition FF is needed, which can 
currently be calculated only with uncertainties in the $15-20$\% range.
All hopes rest, again, on LQCD to eventually provide unquenched 
calculations with much improved precision for channels with stable
particles (\ie, $B\ra \pi\ell\nu$) or {\it narrow} resonances
(\ie, $B\ra \eta/\omega\, \ell\nu$).

\subsection{From Exclusive Fully-leptonic $B$~Decays}

Fully-leptonic $B$~decay provide the theoretically cleanest way to 
determine $|\Vub|$:
\begin{equation}
 \Gamma(B^-\ra\ell^-\nu) = \frac{G_F^2}{8\pi} m_B^3 f_B^2 |\Vub|^2 x_\ell(1-x_\ell)^2,\ {\rm with}\ x_\ell = m_\ell/m_B.
\label{eq:Bellnu}
\end{equation}

Unfortunately, the current best LQCD determination of the $B$~decay constant
comes with a 15\% error: $f_B = (200 \pm 30)\mev$.
With a SM expectation of                   
$B(B^-\ra \tau^-\nu) \simeq 1.1\times10^{-4}$,
observation of this decay is just around the corner, exploiting again 
the Breco technique. Hopefully, LQCD will manage to push down their error
as fast as the $B$~factories accumulate data.

\section{\Vtd, \Vts}

The third side of the UT, $R_t$, is determined by $|\Vtd|$.
It can be extracted from \dmd, the 
$B^0_d - \bar{B}^0_d$ 
mass difference measured in mixing:
\begin{equation}
 \Delta m_d \propto (\sqrt{B_{B_d}} f_{B_d})^2 |\Vtd|^2 \propto R_t^2 = \frac{1}{\lambda^2}\,\frac{|\Vtd|^2}{|\Vcb|^2} = (1 - \bar{\rho})^2 + \bar{\eta}^2,
\label{eq:dmd}
\end{equation}

where LQCD provides~\cite{PDGmix}
$(\sqrt{B_{B_d}} f_{B_d}) = (230 \pm 40)\mev$
with a 17\% uncertainty.

It has been customary to use the {\it ratio}
$\Delta m_d/\Delta m_s$,
where $\Delta m_s$ is the corresponding mass difference from 
$B^0_s - \bar{B}^0_s$ mixing,
because the {\it ratio} 
$\xi = (\sqrt{B_{B_d}} f_{B_d})/(\sqrt{B_{B_s}} f_{B_s})$
was believed to be better known (and most other factors cancel).
Using $\xi = 1.16 \pm 0.05$ 
together with the 
LEP/SLD/CDF mixing amplitude analysis, 
$\Delta m_s > 14.9\ps^{-1}$ at 95\%~CL,
one can get a significant constraint on $R_t$:
\begin{equation}
                 \frac{\Delta m_d}{\Delta m_s} =
                 \frac{m_{B_d}}{m_{B_s}} \xi \frac{|\Vtd|^2}{|\Vts|^2}
                 \propto
                 \frac{1}{\xi^2} \lambda^2 
                 \left[ (1 - \bar{\rho})^2 + \bar{\eta}^2 \right] = 
                 \frac{1}{\xi^2} \lambda^2 R_t^2
\label{eq:dmddms}
\end{equation}

However, it was recently discovered~\cite{Kronfeld} that 
$\xi$ might have to change by 4$\sigma^{old}$ and assume an error twice
as large as currently, resulting in a much relaxed constraint on $R_t$.

\section{Conclusions}

(Inclusive) $|\Vcb|$ is becoming a precision quantity with an uncertainty of 
less than 3\%; it could become even more precise with better measurements 
of HQE non-perturbative parameters -- barring quark-hadron duality violations.

$|\Vub|$ is far from that, with an uncertainty of about 15\%. There are good
prospects, though, for significant improvements over the next few years:
new analysis approaches can significantly reduce both the experimental
{\it and} theoretical errors of the inclusive measurements. 
The exclusive measurements will equally benefit, contingent upon sufficient
progress in LQCD.

$|\Vtd|$ and $|\Vts|$ are awaiting a more precise $\xi$~ratio from LQCD
and, more importantly, a {\it measurement} of (not a {\it limit} on) 
$\Delta m_s$.

$|\Vud|$ will benefit from consistent neutron $\beta$-decay data on 
$\lambda = g_A/g_V$ and from better pion $\beta$-decay data.

$|\Vus|$ is awaiting a (theoretical) conclusion on the right value and 
error for the $K\ra\pi$ transition FF.

\section{Acknowledgements}
Work supported by Department of Energy contract DE-AC03-76SF00515.


\begin{thebibliography}{99}
\bibitem{PDGckm}Particle Data Group, K.~Hagiwara{\it et al},
Phys. Rev. {\bf D66}, 010001-113 (2002) and references therein.
\bibitem{PDGvcb}Particle Data Group, K.~Hagiwara{\it et al},
Phys. Rev. {\bf D66}, 010001-701 (2002) and references therein.
\bibitem{PDGvub}Particle Data Group, K.~Hagiwara{\it et al},
Phys. Rev. {\bf D66}, 010001-706 (2002) and references therein.
\bibitem{PDGsm}Particle Data Group, K.~Hagiwara{\it et al},
Phys. Rev. {\bf D66}, 010001-98 (2002) and references therein.
\bibitem{burasfleischer}
A.J.~Buras, R.~Fleischer, Quark Mixing, CP Violation And Rare Decays After
The Top Quark Discovery,
in: Heavy Flavour II, World Scientific (1997).
\bibitem{lettsmaettig}J.~Letts, P.~M\"attig,
Eur. Phys. J. {\bf C21}, 211 (2001).
\bibitem{HF9bauer}C.~Bauer, hep-ph/0112243.
\bibitem{LEPvcb}LEP \Vcb\ Working Group, Internal Note,
\url{http://lepvcb.web.cern.ch/LEPVCB/}.
\bibitem{CLEOinclVcb}D.~Cronin-Hennessy {\it et al},
Phys. Rev. Lett. {\bf 87}, 251808 (2001).
\bibitem{Bigi9905}I.~Bigi, hep-ph/9907270.
\bibitem{CERNep2001050}D.~Abbaneo \etal, CERN-EP/2001-050 and hep-ex/0112028.
\bibitem{CLEOlepmom}R.A.~Briere \etal, hep-ex/0209024.
\bibitem{DELPHIlepmom}M.~Battaglia, M.~Calvi, L.~Salmi, DELPHI 2002-071-CONF-605.
\bibitem{BABARichepMX}B.~Aubert \etal, hep-ex/0207084.
\bibitem{LLW}A.K.~Leibovich, Z.~Ligeti, M.B.~Wise, hep-ph/0205148.
\bibitem{BLM}C.~Bauer, M.~Luke, T.~Mannel, hep-ph/0205150.
\bibitem{Neubert}M.~Neubert, hep-ph/0207002.
\bibitem{Bornheim}A.~Bornheim \etal,
Phys. Rev. Lett. {\bf 88}, 231803 (2002).
\bibitem{BABARichepEll}B.~Aubert \etal, hep-ex/0207081.
\bibitem{Kronfeld}A.~Kronfeld, these proceedings and hep-ph/0209231.
\bibitem{PDGmix}Particle Data Group, K.~Hagiwara{\it et al},
Phys. Rev. {\bf D66}, 010001-670 (2002) and references therein.
\end{thebibliography}
\end{document}